\documentclass[aps,pre,twocolumn,reprint,tightenlines,
amsmath,amssymb,amsfonts,showpacs,showkeys,floatfix]{revtex4-1}
\usepackage{graphicx}
\usepackage{dcolumn}
\usepackage{bm}
\usepackage{enumerate}
\usepackage{array}
\usepackage{enumitem}
\usepackage{color}
\usepackage{braket}
\usepackage{csquotes}
\newcommand{\beq}{\begin{equation}}
\newcommand{\eeq}{\end{equation}}
\begin{document}
\title{Efficient Density Matrix Renormalization Group algorithm to study Y-Junctions 
with integer and half-integer spin}
\author{Manoranjan Kumar$^1$, Aslam Parvej$^{1}$, Simil Thomas$^2$, S. Ramasesha$^3$ and Z. G. Soos$^4$}
\email{manoranjan.kumar@bose.res.in, aslam12@bose.res.in,
simil.thomas@kaust.edu.sa, ramasesh@sscu.iisc.ernet.in, 
soos@princeton.edu}

\affiliation{$^1$S. N. Bose National Centre for Basic Sciences, Calcutta, Calcutta 700098, India \\
$^2$ Solar \& Photovoltaics Engineering Research Center, King Abdullah University of Science and 
Technology, Thuwal 23955-6900, Kingdom of Saudi Arabia \\
$^3$Solid State and Structural Chemistry Unit, Indian Institute of Science, Bangalore 560012, India \\
$^4$Department of Chemistry, Princeton University, Princeton, New Jersey 08544, USA}
\date{\today}
\begin{abstract}
An efficient density matrix renormalization group (DMRG) algorithm is presented and applied to Y-junctions, systems 
with three arms of $n$ sites that meet at a central site. The accuracy is comparable to DMRG of chains. As in chains, 
new sites are always bonded to the most recently added sites and the superblock Hamiltonian contains only new or 
once renormalized operators. Junctions of up to $N = 3n + 1 \approx 500$ sites are studied with antiferromagnetic 
(AF) Heisenberg exchange $J$ between nearest-neighbor spins $S$ or electron transfer $t$ between nearest neighbors 
in half-filled Hubbard models. Exchange or electron transfer is exclusively between sites in two sublattices with
$N_A \ne N_B$. The ground state (GS) and spin densities $ \rho_r =  <S^z_r >$ at site $r$ are quite different for 
junctions with $S$ = 1/2, 1, 3/2 and 2. The GS has finite total spin $S_G = 2S (S)$ for even (odd) $N$ and for 
$M_G =S_G$ in the $S_G$ spin manifold, $\rho_r > 0 (< 0)$ at sites of the larger (smaller) sublattice. $S$ = 1/2 
junctions have delocalized states and decreasing spin densities with increasing $N$. $S$ = 1 junctions have four 
localized $S_z = 1/2$ states at the end of each arm and centered on the junction, consistent with localized states 
in $S$ = 1 chains with finite Haldane gap. The GS of $S$ = 3/2 or 2 junctions of up to 500 spins is a spin density 
wave (SDW) with increased amplitude at the ends of arms or near the junction. Quantum fluctuations completely suppress 
AF order in $S$ = 1/2 or 1 junctions, as well as in half-filled Hubbard junctions, but reduce rather than suppress 
AF order in $S$ = 3/2 or 2 junctions.
\end{abstract}
\maketitle
\section{Introduction}
\label{Sec:I}
The transport and magnetic properties of a system with a junction of three wires have been a 
frontier area of research. Y junctions such as 3-terminal Josephson devices \cite{1} or carbon 
nanotubes \cite{2} have been studied experimentally. Understanding quantum effect in three terminal 
junctions is important for potential applications as rectifiers \cite{2,3}, switches and logic gate 
devices \cite{4}. Recently these systems have also been studied theoretically. \cite{5,6,7,8,9,10,11,12} 
Interesting predictions include a low energy chiral fixed point with an asymmetric 
current flow in a spinless fermionic system \cite{13} and negative density reflection at the 
junction of Bose liquid of ultra-cold atoms \cite{14}. Theoretical studies have been mainly based 
on field theoretical approaches \cite{14,15}.\\
 
Exact numerical results are limited to very small junctions and tend to be inconclusive, especially 
with respect to quantum many body effects. Numerical techniques such as density matrix renormalization 
group (DMRG) give excellent results for one-dimensional (1D) systems \cite{16}. At first sight, however, 
DMRG appears to be far less accurate for structures with three terminals in which a long bond is repeatedly 
renormalized. Guo and White (GW) introduced \cite{17} a new DMRG algorithm, summarized in Section {{II}}, 
for Y junctions with spin $S$ at every site. We present in this paper a modified DMRG algorithm, quite distinct 
from that of GW, for Y-junctions with $N=3n+1$ sites and three equal 1D arms of $n$ sites. The accuracy
and efficiency of the modified algorithm is comparable to DMRG in 1D chains, and we have studied Y junctions
of up to 500 spins. We note that tensor-tree networks \cite{17a} are a general approach to many-body systems with a tree
structure such as Y junctions, dendrimers or Bethe lattices. Stilbenoid dendrimers are a recent quantum chemical
application \cite{17b} based on molecular units with many degrees of freedom. Tree networks based on different units call
for diverse algorithms. \\

\begin{figure}[t]
\begin{center}
\includegraphics[width=0.55\textwidth]{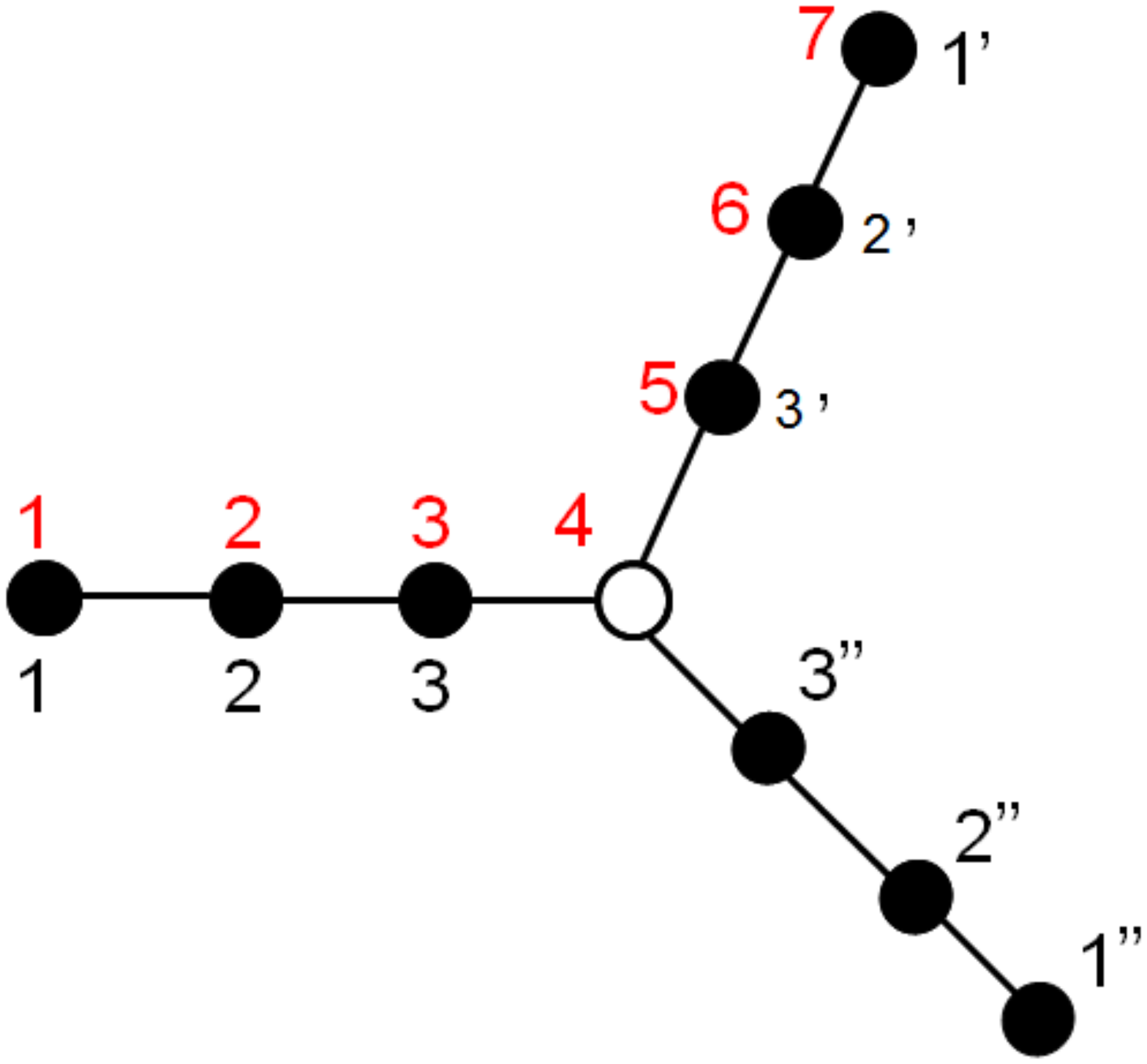}
\caption{Y junction of $N=10$ sites with three equal arms of $n=3$ sites. The numbering for the 
left (unprimed), up (primed) and down blocks (double primed) is used in the DMRG algorithm; the 
numbering in red is used for spin densities.}
\label{Fig1}
\end{center}
\end{figure}

We consider Y junctions with antiferromagnetic Heisenberg exchange $J$ between sites with spin $S$. The 
Hamiltonian for the junction in Fig. \ref{Fig1} is 
\begin{equation}
H_S=J\sum_{\langle{rr'} \rangle}{\bf{s}}_{r} \cdot {\bf{s}}_{r'}.
\label{eq1}
\end{equation}
\noindent
The sum is restricted to adjacent sites and $J=1$ is 
a convenient unit of energy. We discuss systems with $S=1/2$, 1, 3/2 or 2 and  
also study fermionic junctions that correspond 
to half-filled Hubbard models with $N$ electrons and $N$ sites,
\begin{equation}
H_F=-t\sum_{{\langle{pp'}\rangle}\sigma}(a_{p\sigma}^{+}a_{{p'}\sigma}
+a_{{p'}\sigma}^{+}a_{p\sigma})+U\sum_{p}a_{p\alpha}^{+}a_{p\beta}^{+}a_{p\beta}a_{p\alpha}.
\label{eq2}
\end{equation}
\noindent
Electron transfer $t$ (set to 1 to define the energy scale) is limited to adjacent sites $p,{p'}$, 
the number operator is $n_p$, and 
$U>0$ is repulsion for two electrons at a site. The H\"{u}ckel or tight-binding limit of $U=0$ 
is readily solved exactly and provides a direct check of accuracy. Trimethylenemethane, $C_4H_6$, 
is a Y junction with four $\pi$-electrons, four C atoms and has a stable triplet GS \cite{18}. 
Bipartite H\"{u}ckel models with many singly occupied $\pi$-orbitals are a design 
principle for high spin hydrocarbons \cite{19}. The atomic limit $U>>t$ reduces $H_F$ at half 
filling to $H_S$ with $S=1/2$ and $J=4t^2/U$.\\
 
Both $H_S$ and $H_F$ conserve total spin $S_T$ and its component $S_z$. By convention, we choose 
the Zeeman component $S_z=S_T$ when the ground state (GS) has spin $S_G>0$. The spin density at 
site $r$ is the GS expectation value
\begin{equation}
{\rho_r}=\langle{S_r^z}\rangle
\label{eq3}
\end{equation}
The sum over sites $r$ returns $S_z\geq0$, but individual $\rho_{r} $ may be positive or negative. 
Y junctions are bipartite: All exchange $J$ or electron transfer $t$ is between sites that form two 
sublattices, A and B, here with $N_A\ne N_B$ sites. The GS of $H_S$ has $S_G=S|N_A-N_B|$, which 
alternates between $S_G=S$ and $2S$ for odd and even $N$, respectively. Sites in the larger 
sublattice have positive $\rho_{r}$, those in the smaller sublattice have negative $\rho_{r}$.\\

The paper is organized as follows. The modified algorithm for Y junctions with equal arms is
presented and tested in Section \ref{Sec:II}, including both infinite and finite DMRG algorithms. Its accuracy is 
fully comparable to DMRG for chains. As in chains, new sites are always bonded to the most recently 
added sites and the superblock Hamiltonian contains only new or once renormalized operators. 
The algorithm is applied to Y junctions in Section \ref{Sec:III}, first to fermionic and $S=1/2$ 
junctions, then to $S=1$ junctions and finally to $S=3/2$ and 2 junctions. We focus on spin 
densities and size dependence. Localized states in $S=1$ junction are in excellent agreement with 
the valence bond solid (VBS) model of Affleck, Kennedy, Lieb and Tasaki (AKLT) \cite{20}. There is 
a localized $S_z=1/2$ state at the end of each arm and one centered on the junction. The localization 
length $\xi=6.25$ in arms is close to the chain result of White and Huse \cite{21} while the length 
$\xi_J<\xi$ indicates greater localization at the junction. The GS of $S=3/2$ or 2 junctions up to  
500 spins are unexpectedly different, however: They are spin density waves (SDWs) with increased amplitude at 
the ends of arms and near the junction. Antiferromagnetic (AF) order is possible in systems with 
$S_G>0$ and found in Y junctions of $S > 1$ spins. Quantum fluctuations suppress AF completely in 
$S=1/2$ or 1 junctions with long arms, but only partially in $S > 1$ junctions. We briefly mention 
in the Discussion generalizations of the algorithm to other junctions.

\section{Modified DMRG Algorithm}
\label{Sec:II}
The computational effort for one eigenstate in conventional DMRG for 1D chains with open boundary 
conditions goes as $O(Nm^4)$ where $N$ is the number of sites and $m$ is the number of states per 
block for a given accuracy \cite{22,23}. The reason why is as follows: The number of arithmetic 
operations to obtain all eigenvalues of an $L \times L$ matrix is $O(L^3)$; so the number of operations 
for one eigenvalue goes as $L^2$. In DMRG the matrix size is $L=16m^2$ for fermions and $L=(2S+1)^2m^2$ 
for spin $S$. In either case, $L^2$ goes as $m^4$ and a system of size $N$ requires DMRG steps of 
$N/2$. This estimate excludes the construction and diagonalization of density matrices which are $O(m^3)$. 
The greatest cost is the superblock diagonalization that goes as $O(m^4)$ for one eigenstate.\\

Conventional DMRG for Y junctions scales as $O(m^6)$ and, as shown in Fig. \ref{Fig1}c of ref. 17, involves 
a long bond whose operators are renormalized many times. GW \cite{17} cite previous DMRG applications 
to Y junctions and present a more efficient scheme for junctions with three equal arms that meet at 
a point, as in Fig. \ref{Fig1}, or at the vertices of an equilateral triangle. Singular 
value decomposition is used to obtain the density matrix of a single arm, and is faster as it requires $O(m^4)$ 
operations instead of $O(m^6)$. The method has a large truncation of the density matrix when two 
arms are combined into a single block. We avoid truncation below since arms are never combined. 
The slow $O(m^6)$  step is the GS eigenvector of the superblock matrix, which goes as $L = O((2S+1)m^3)$ 
for three arms. The modified algorithm has improved accuracy for smaller $m$ and makes it possible to 
treat Y junctions of $N\sim 500$ sites.\\

DMRG algorithms of 1D chains add two sites per step between the left and right blocks.\cite{22,23} 
The new sites are always bonded to the most recently added sites. The superblock Hamiltonian of 
chains only contains new operators or once renormalized operators, a desirable featured that we 
retain for Y junctions. In some algorithms the number of newly added site in the superblock can 
vary from one \cite{24}, two \cite{17} or four \cite{5} depending on the accuracy requirements 
of the systems. Here the superblock grows by three sites.\\ 

\begin{figure}[b]
\begin{center}
\includegraphics[width=0.55\textwidth]{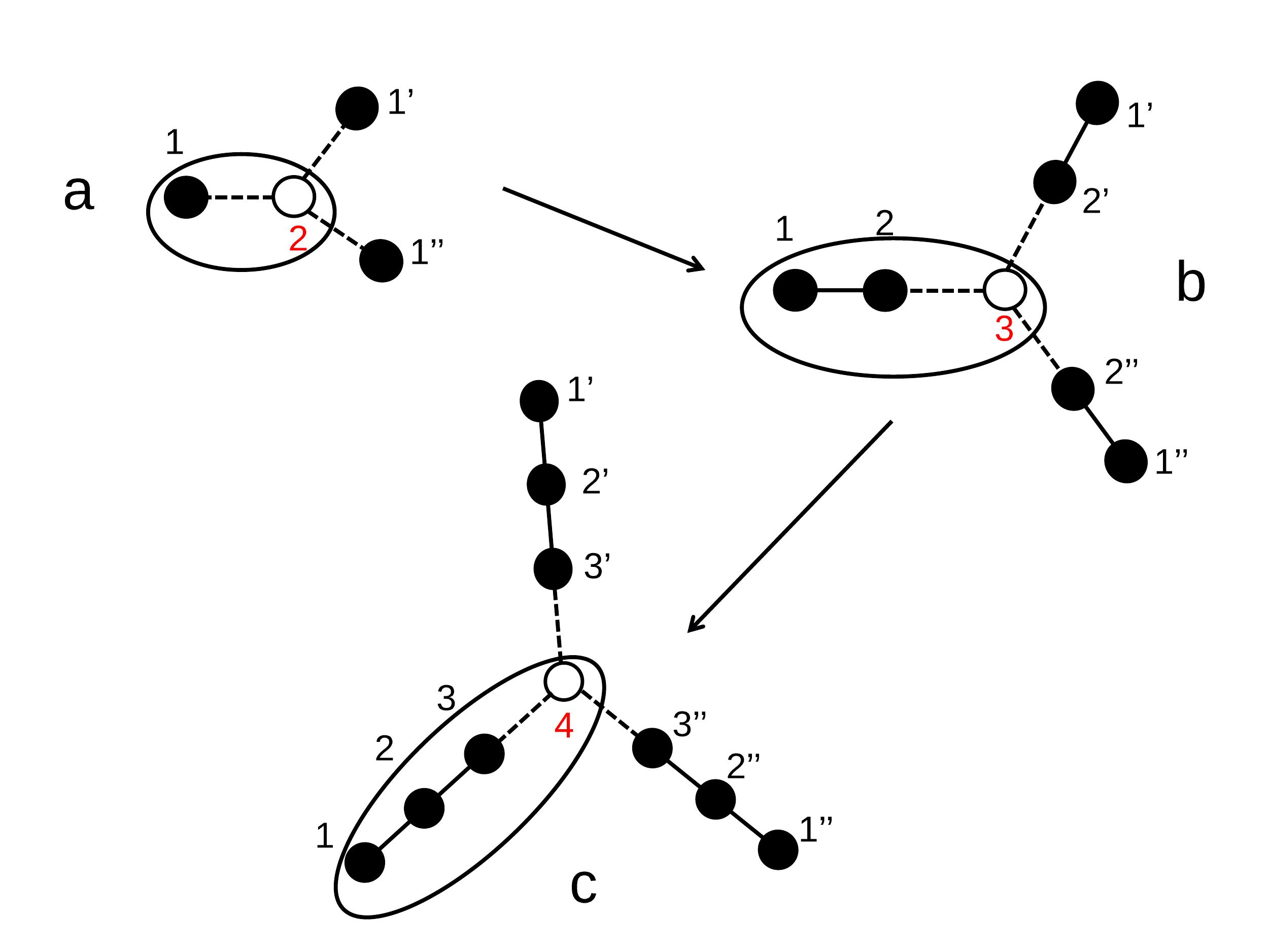}
\caption{Schematic representation of the infinite DMRG algorithm for Y junctions with equal arms. 
At each step, the loop encloses the system and the superblock contains a new site shown as an open dot,
 and three arms.}
\label{Fig2}
\end{center}
\end{figure}

The modified algorithm is shown schematically in Fig. \ref{Fig2}. Sites enclosed in loops define systems that 
consist of an arm plus a new site. The environment contains all sites in the other two arms. 
The key point is that the system at one step becomes an arm in the next step, thereby avoiding 
having to combine two arms into one block. The system block has $m \times r$ degrees of freedom, 
$m$ for the arm and $r$ for the new site, with $r=4$ for fermionic junctions and $r=2S+1$ for sites with 
spin $S$. The relevant dimension is not $mr$ of the system block, however, because the density 
matrix is block diagonal in sectors with different $M_S$ values. The time needed to diagonalize 
the density matrix is negligible in sectors of dimension $mr/(NS+1)$. We obtain all 
density matrix eigenvectors $\vert{i}\rangle$ of the system block by block 
diagonalizing it into different $M_S$ sectors.\\

As shown in Fig. \ref{Fig2}, we start with a superblock of four sites, $N=3n+1=4$. The notation $(2, 1, 1)$ 
refers to an arm plus the central site and two other arms, respectively. The second step corresponds to 
$(3, 2, 2)$ and $N=7$, the third to $(4, 3, 3)$ and $N=10$, and so on. The $n+1$ sites of arm plus central site at 
step $n$ become the arm at step $n+1$. We find the GS eigenvector $\vert{\psi}\rangle$ of the superblock, 
starting with $N=4$, and expand $\vert{\psi}\rangle$ in the basis of the system (arm plus site) and the 
environment (two arms),

\begin{equation} 
{\vert{\psi}\rangle}=\sum_{ik}C_{ik}\vert{i}\rangle\vert{k}\rangle.
\label{eq4}
\end{equation} 
\noindent
The basis vectors $\vert{k}\rangle$ are direct products of basis states of two arms of the system of the previous step. 
The total number of sites, $N=3n+1$, increases by three at each step. The reduced density matrix 
of the system has elements

\begin{equation} 
{\rho_{ij}}=\sum_{k}C_{ik}^{*}C_{jk}.
\label{eq5}
\end{equation} 
\noindent
The sum is over the environmental degrees of freedom. We suppose $\rho$ to be a matrix of dimension $M$. 
After diagonalization, we take the $m$ eigenvectors of $\rho$ with the largest eigenvalues as elements 
of an $M{\times}m$ matrix $\rho'$. The effective Hamiltonian and operators in the truncated 
$m{\times}m$ basis are renormalized according to

\begin{equation}
\begin{aligned}
O &= (\rho')^{\dagger}O{\rho'}\\
H &= (\rho')^{\dagger}H{\rho'}
\end{aligned}
\label{eq6}
\end{equation} 

\noindent
where $(\rho')^{\dagger}$ is the transpose matrix, and $O$ and $H$ are the operators and Hamiltonian 
of system block. The superblock eigenvalue calculation is the slow step that scales as $O(m^6)$, although 
conservation of total $S_z$ reduces the dimension to less than $(2S+1)m^3$. The GS then yields the reduced 
density matrix $\rho_{ij}$ of the system for the junction in which each arm is one site longer. Since 
operators of the system block are renormalized only once, similar to 1D chains, we expect similar 
truncation errors in Y junctions.\\ 
 
The following steps and Fig. \ref{Fig2} describe the infinite DMRG algorithm for Y junctions with equal arms:

\begin{enumerate}[label=(\alph*)]
\item{Start with four sites, the superblock in Fig. \ref{Fig2}a.}
\item{Find the GS eigenvalue and eigenvector.}
\item{Construct the density matrix of the system block, shown in Figs \ref{Fig2}a, \ref{Fig2}b and \ref{Fig2}c 
for 4,7 and 10 site superblocks, respectively. Diagonalize it to get the eigenvectors corresponding to 
the $m$ largest eigenvalues.}
\item{Renormalize the operators and Hamiltonian for the system blocks using Eq. \ref{eq6}.}
\item{Construct the Hamiltonian of the superblock as shown Fig. \ref{Fig2}b and \ref{Fig2}c.}
\item{Repeat the process from b to e until the desired size $N=3n+1$ is reached.}
\end{enumerate}

\begin{figure}[b]
\begin{center}
\includegraphics[width=0.5\textwidth]{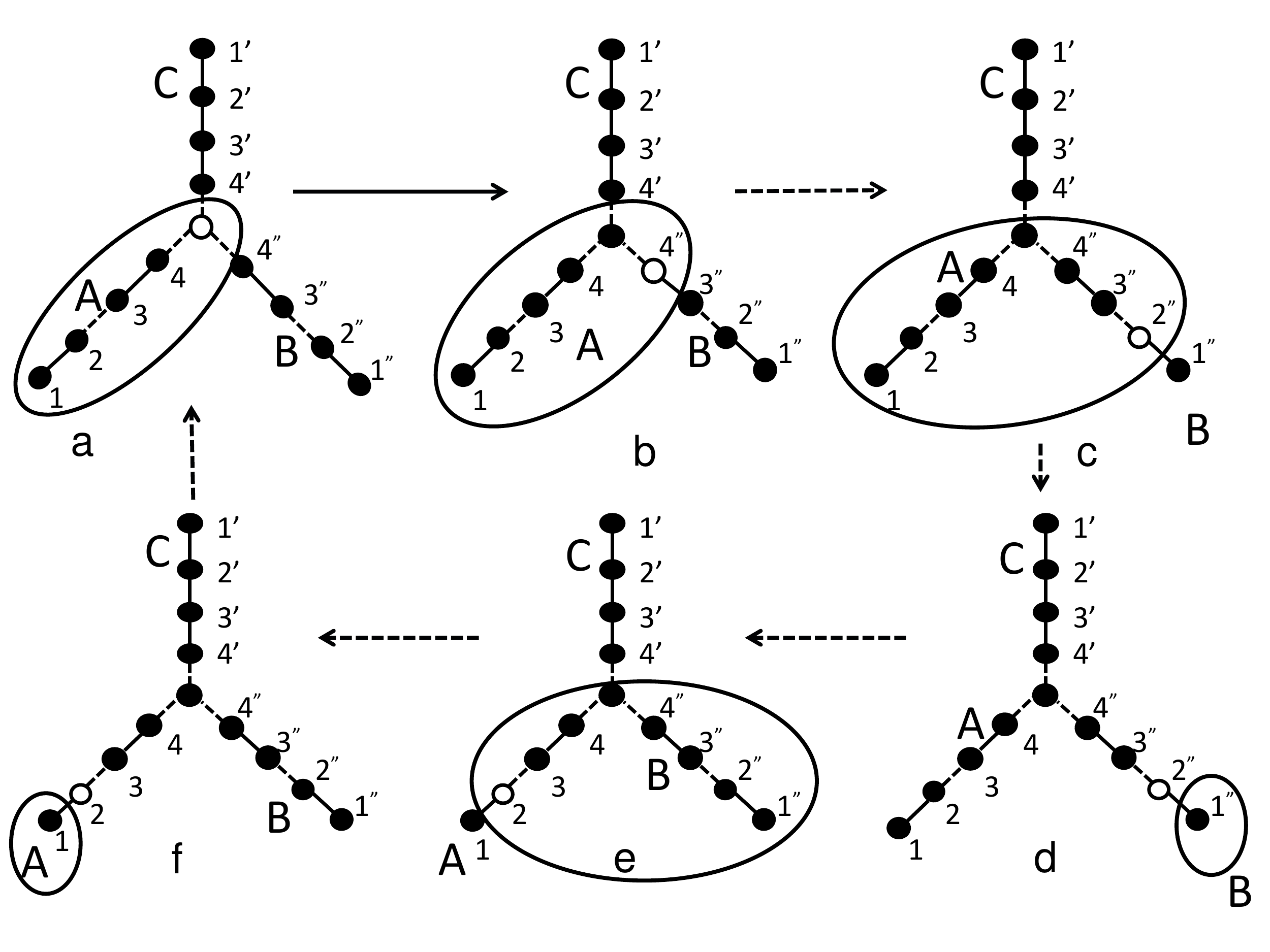}
\caption{Schematic representation of finite DMRG steps. Sites of the system block are 
enclosed in the loop; remaining sites are in the environmental block.The new site 
is the open dot. A, B and C arm refer to the three different arms.} 
\label{Fig3}
\end{center}
\end{figure}

Finite DMRG is required to obtain accurate spin densities, correlation functions and other GS 
properties. The conventional finite algorithm for 1D chains has two new sites and sweeps through 
two arms of the same chain \cite{22,23}. The algorithm for Y junctions has one new site and sweeps 
through two arms while keeping the third arm constant. The procedure is shown schematically in 
Fig. \ref{Fig3} and summarized below. Three to four DMRG sweeps are typically sufficient for 
converged energies. Finite DMRG is particularly important for junctions with $S=3/2$ and 2 sites.

\begin{enumerate}[label=(\alph*)]
\item{Start with the superblock with equal arms as shown Fig. \ref{Fig3}a taken from the infinite 
algorithm calculation. Select two arms A and B for sweeping through.}
\item{Find the GS eigenvector of the superblock. The new system block \enquote*{A} is the old block A plus 
a new site, shown as the open dot in Fig. \ref{Fig3}b. Block \enquote*{A} has one site added to arm A and 
removed from arm B at every step while arm C remains same. Construct the reduced density matrix of 
the system block \enquote*{A}.}
\item{Renormalize the Hamiltonian and operators of the system block \enquote*{A}.}
\item{Construct the superblock Hamiltonian with new blocks \enquote*{A}, \enquote*{B} and \enquote*{C} shown 
in Fig. \ref{Fig3}b.}
\item{Repeat the steps b to d until block \enquote*{B} has only one site as shown in Fig. \ref{Fig3}c.}
\item{Now take \enquote*{B} as the system block, block \enquote*{A} and \enquote*{C} as the environment. As shown 
in Fig. \ref{Fig3}d, add a site to block \enquote*{B} and remove one from block \enquote*{A}.}
\item{Repeat steps b to d for system block \enquote*{B} until block \enquote*{A} has only one site as shown 
in Fig. \ref{Fig3}e.} \item{In the next step, \enquote*{A} becomes the system block. One site is added in 
\enquote*{A} and removed from block \enquote*{B}. Steps b to d are repeated until Fig. \ref{Fig3}a is reached.}
\item{Now take blocks \enquote*{A} and \enquote*{C} while keeping block \enquote*{B} constant and repeat the 
cycle a to h that starts and ends with equal arms in Fig. \ref{Fig3}a. Finally, take block \enquote*{B} and 
\enquote*{C} while keeping the block \enquote*{A} constant. Repeat the cycle from a to h that starts and ends
with Fig. \ref{Fig3}a.}
\item{One cycle of finite DMRG is the whole process from a to i.}
\end{enumerate}

\begin{table}[t]
\centering
\caption{The $m$ dependence of the GS energy per site $\varepsilon_{0}$ of Y junctions with 
64 sites, equal arms, and $U=0$ in Eq. \ref{eq2} or $S=1/2$, $J=1$ in Eq. \ref{eq1}.}
\begin{tabular}{|c |>{\centering\arraybackslash} m{10em} |>{\centering\arraybackslash} m{10em} |} 
\hline
$m$ & $U=0$ & $S=1/2$ \\ 
\hline
 20  & -1.23336877828  & -0.43915791387  \\[1ex] 
 40  & -1.25809972580  & -0.43915891861 \\[1ex]
 60  & -1.25826370430 &  -0.43915892503 \\[1ex]
 80  & -1.25838125000 &  -0.43915892523 \\[1ex]
 100 & -1.25842968750 &  -0.43915892525 \\ [1ex]
 Exact & -1.25848468281 &        -       \\ [1ex] 
\hline
\end{tabular}
\label{table1}
\end{table}

Next we discuss the accuracy and efficiency of the algorithm. The $U=0$ limit of Eq. \ref{eq2} is a 
H\"{u}ckel or tight-binding model of non-interacting electrons on $N$ sites that can readily be solved 
exactly. As an example, we took a half-filled band of $N=64$ sites and calculated the GS energy per 
site $\varepsilon_{0}$ as a function of $m$, the dimension of the system block in the truncated basis. 
Table \ref{table1} shows good convergence by $m \sim 60$ for this fermionic system of about $4^{N}$ 
degrees of freedom, or some $4^{21}$ per arm. DMRG of non-interacting electrons often converges the most 
slowly due to GS degeneracy or to higher entanglement entropy \cite{24,26}. The Y junction of $S=1/2$ 
spins in Eq. \ref{eq1} is the $U \gg t$ limit with $2^{N}$ spin degrees of freedom whose GS energy per 
site is not known exactly. As shown in Table \ref{table1}, $m \sim 20$ is sufficient for  $\varepsilon_{0}$ 
of junctions with 21 spins per arm.\\

\begin{table}[b]
\centering
\caption{Truncation errors $P(m)$ of 64-site Y junction of $S = 3/2$ and 2 as a function of $m$.}
\begin{tabular}{|c |>{\centering\arraybackslash} m{12em} |>{\centering\arraybackslash} m{12em} |}
\hline
$m$ & $P(m), S=3/2$ & $P(m), S=2$ \\
\hline
 64  & $1.2\times10^{-9}$  & $7.2\times10^{-6}$  \\[1ex]
 80  & $7.6\times10^{-10}$ & $3.0\times10^{-6}$   \\[1ex]
 100 & $1.7\times10^{-10}$ & $1.5\times10^{-6}$    \\[1ex]
 130 & $5.2\times10^{-11}$ & $5.6\times10^{-7}$    \\ [1ex]
\hline
\end{tabular}
\label{table2}
\end{table}

As additional tests of the algorithm, we consider the total energy $E(m)$ of 64-site Y junctions with $J=1$ in 
Eq. \ref{eq1} and $S=1/2$, 1, 3/2 and 2 as a function of $m$. The truncation errors $P(m)=1-\sum_{j}^{m}\lambda_{j}$ 
on keeping $m$ eigenvalues of the density matrix are listed in Table \ref{table2} for $S = 3/2$ and 2. 

\begin{figure}[t]
\begin{center}
\includegraphics[width=0.5\textwidth]{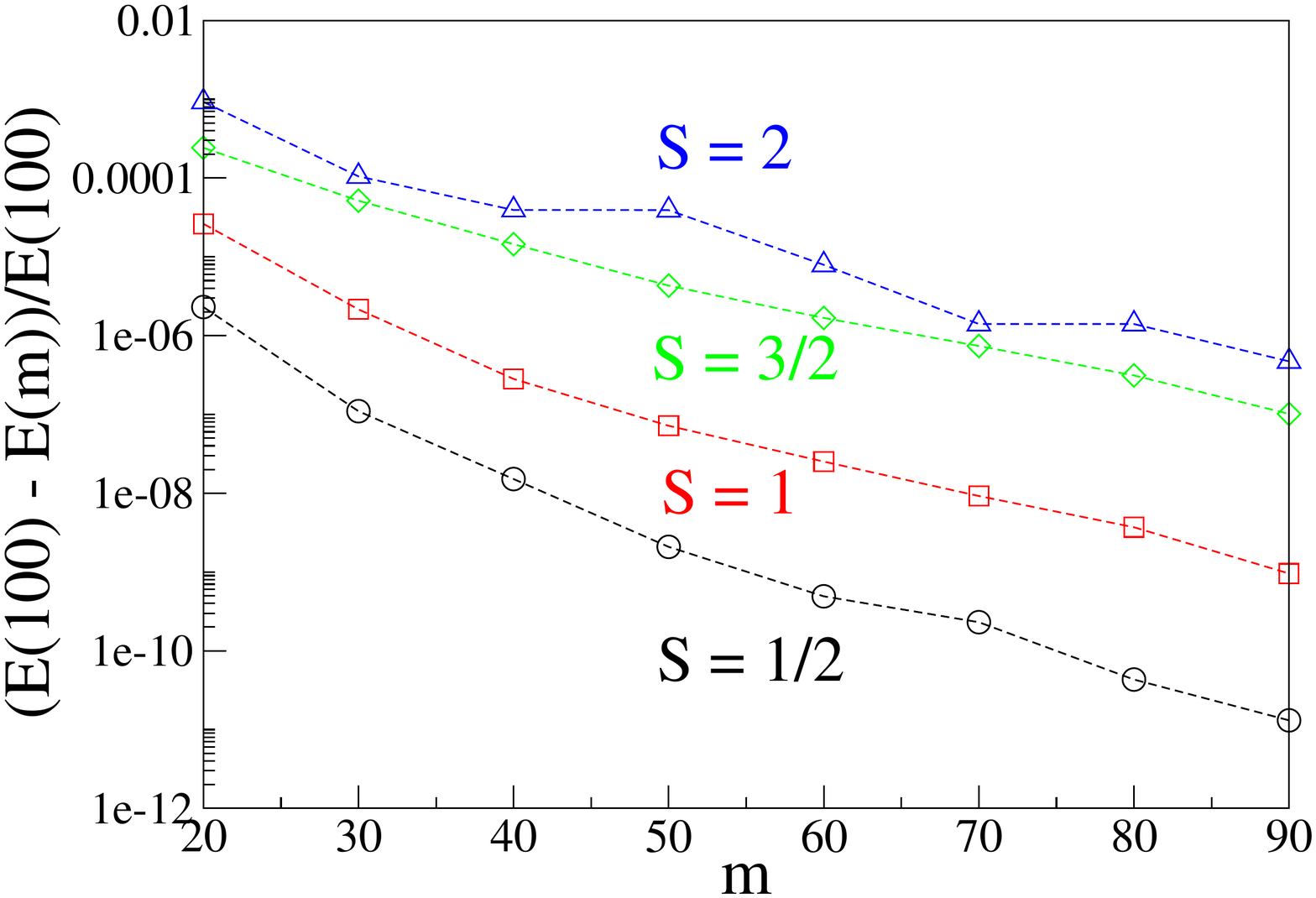}
\caption{Total GS energy $E(100)-E(m)$ as a function of $m$ for 64-site Y junctions with 
equal arms, $J=1$, and the indicated $S$ per site in Eq. \ref{eq1}.}
\label{Fig4}
\end{center}
\end{figure}

Since the 
exact GS is not known, we follow the evolution of $\Delta{{E(m)}}=E(m_{0})-E(m)$ where $m_{0}=100$ is the nominally 
the converged value. Excellent convergence is achieved in Fig. \ref{Fig4} by $m \sim 70$, with $\Delta{E}$ of the 
order of $10^{-10}$ for $S=1/2, 10^{-7}$ for $S=1$, and $10^{-6}$ for $S=3/2$ or 2. Increasing $m$ to 130 lowers 
$\Delta{E/E(100)}$ by $5 \times 10^{-7}$ for $S = 2$. The $P(m)$ change in Table \ref{table2} is also small. 
By contrast, the GW algorithm \cite{17} for $\Delta{E}$ with $S=1$ reaches only $10^{-6}$ around $m=140$ in Fig. \ref{Fig3} of ref. 17. The present algorithm is well suited for Y junctions, both because as in 1D chains operators are renormalized only 
once and because the procedure in Fig. \ref{Fig2} increases the number of sites smoothly without ever having to
combine two arms. 

\section{Localized states and antiferromagnetic order}
\label{Sec:III}
We apply the modified DMRG algorithm to Y junctions with equal arms, either half-filled junctions 
in Eq. \ref{eq2} or Heisenberg junctions with spin $S$ at every site in Eq. \ref{eq1}. Unless 
otherwise stated, the results are based on $m=100$ and $5-10$ sweeps of finite DMRG. We discuss 
junctions of $N=3n+1$ sites, distinguish between odd and even $N$, and study the size dependence. 
The algorithm is applicable to junctions of $N \sim 500$ sites. We focus on AF order in Heisenberg 
junctions with $S > 1$ and on localized states of junctions with integer $S$.\\

As mentioned in the Introduction, Y junctions are bipartite, with different number of sites 
$N_A \ne N_B$ in sublattices A and B. We take $N_A > N_B$ and have

\begin{equation}
\begin{aligned}
N_A &=\frac{N+1}{2}=N_B+1  & & (\text{odd $N$}, \text{even $n$})\\
N_A &=\frac{N}{2}+1=N_B+2  & & (\text{even $N$}, \text{odd $n$}).
\end{aligned}
\label{eq7}
\end{equation}

\noindent
The junction is in sublattice A for odd $N$, in sublattice B for even $N$. The N\'{e}el state 
$\vert {AF}\rangle$ has spins $\pm{S}$ at all sites in sublattices A and B, respectively, and 
is the SDW with the largest possible amplitude; $\vert{AF}\rangle$ is exact in the limit of classical 
spins, $S\rightarrow \infty$. Quantum fluctuations in Eq. \ref{eq1} strongly 
reduce AF order for $S > 1$ and suppress it altogether for $S=1/2$ or 1. Nevertheless, 
$\vert{AF}\rangle$ gives the correct spin, $S_G=2S$ for even $N$ and $S$ for odd $N$, and also 
accounts for the sign of the GS spin densities in Eq. \ref{eq3}, with $\rho_{r} > 0$ for $r$ in 
sublattice A and $\rho_{r} \le 0$ for $r$ in sublattice B. 

\subsection{Fermionic and $S=1/2$ junctions}
The H\"{u}ckel junction has $U=0$ in Eq. \ref{eq2} and $N_A-N_B$ nonbonding orbitals with energy
$\varepsilon=0$ and nodes at all sites in sublattice B. The nonbonding orbitals are easily found 
analytically. The half-filled junction has $N$ electrons, $N$ sites and spin $\alpha$ in nonbonding 
orbitals. The GS for odd $N$ has $S_G=1/2$ and $\rho_{r}=1/(2N_A)$ at sites in sublattice A, 
$\rho_{r}=0$ at sites in sublattice B. The triplet GS for even $N$ has $S_G=1$, $\rho_{r}=1/N_A$ at sites 
in $N_A$ and $\rho_{r}=0$ at sites in $N_B$. Since $S_G=1$ for arbitrarily large (even) $N$, the 
H\"{u}ckel densities at sites in sublattice A decrease as $2/(N + 2)$.\\

\begin{figure}[t]
\begin{center}
\includegraphics[width=0.5\textwidth]{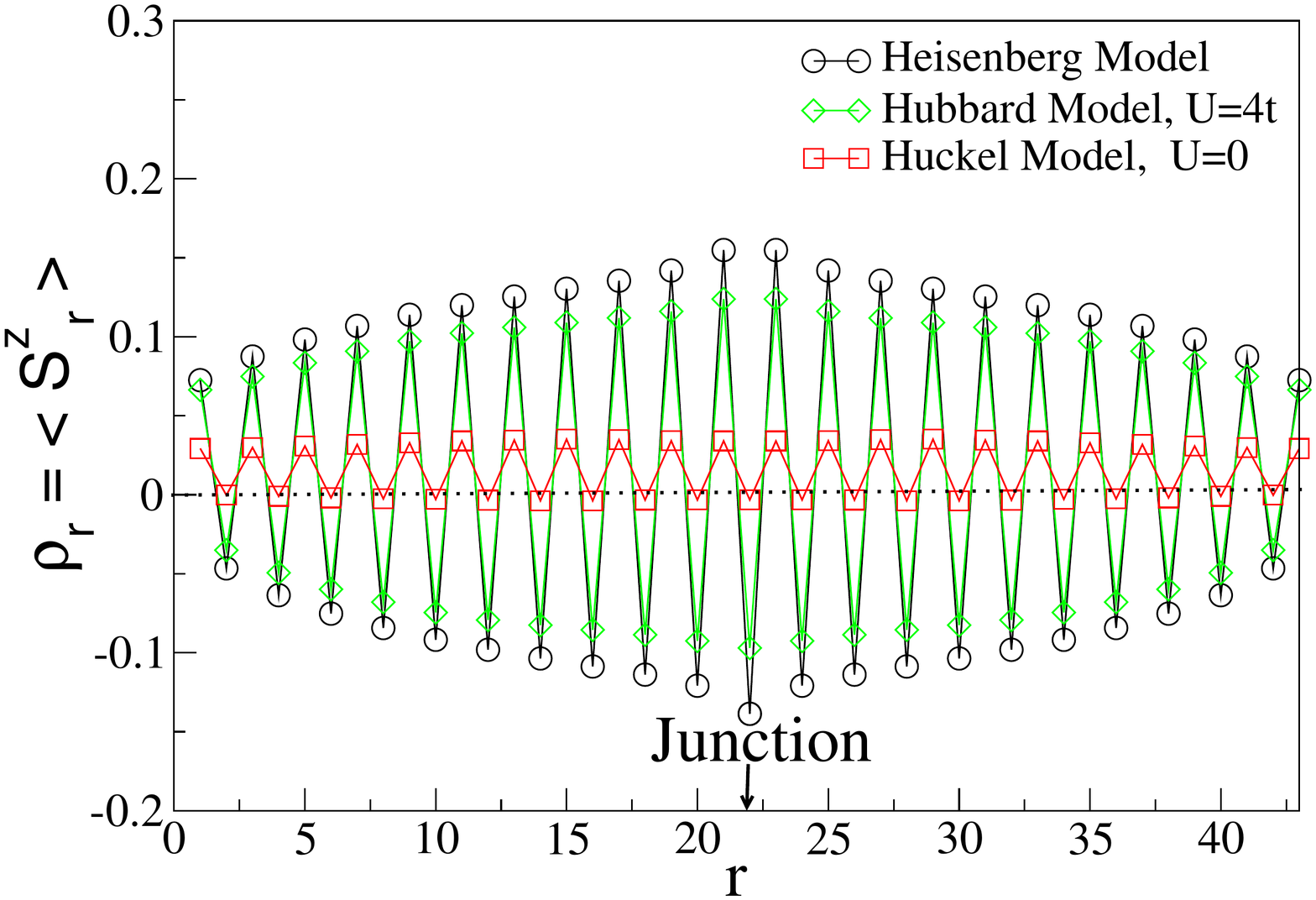}
\caption{Spin densities $\rho_{r}$ along any two arms of 64-site Y junctions with $U=0$ or $4t$ 
in Eq. \ref{eq2}, or $S=1/2$ in Eq. \ref{eq1}. The junction is at $r=22$ with ${\rho_{J}} \leq 0$.} 
\label{Fig5}
\end{center}
\end{figure}

Increasing $U > 0$ in the half-filled junction does not change $S_G$ but induces negative $\rho_{r} < 0$ 
at $N_B$ sites and increases $\rho_{r} > 0$ at $N_A$ sites. The sum over $\vert{\rho_{r}}\vert$ increases 
with $U$ as localized spins are formed due to electron correlations. The spin densities are no longer equal, 
however, as seen in Fig. \ref{Fig5} at $U=4t$, the bandwidth of the 1D H\"{u}ckel or tight-binding model. 
The Heisenberg model with $S=1/2$ in Eq. \ref{eq1} has the largest positive and negative spin densities.\\ 

The Heisenberg junction with even $N$ has $\rho_{J} < 0$ at the junction and spin densities that go as $1/N$. 
The $S=1/2$ junction of $N=202$ spins also has a triplet GS and a spin density distribution similar to 
the Heisenberg model in Fig. \ref{Fig5}. Longer arms lead to smaller spin densities: $\rho_{68}=-0.0929$ 
for the junction at $r=68$; $\rho_{67}=0.0974$ and $\rho_{66}=-0.0817$ at the first and second neighbors 
of the junction; $\rho_{1}=0.0307$ and $\rho_{2}=-0.0198$ at the first two sites of arms. As expected, 
quantum fluctuations entirely suppress AF order in the infinite $S=1/2$ junction. We note that spin 
densities increase along the arms at odd $r$ and become more negative at even $r$. We will later find a 
different pattern in $S=3/2$ junctions in which quantum fluctuations are not as dominant.

\subsection{$S=1$ junctions}
Haldane \cite{27} predicted finite energy gaps $\Delta{(S)}$ in infinite Heisenberg spin chains with 
integer $S$ and nearest neighbor $J>0$. Experimental realizations of $S=1$ chains have confirmed a 
gap that DMRG evaluates \cite{21} as $\Delta{(1})/J=0.4105$. The valence bond solid (VBS) picture of 
AKLT \cite{20} has been widely applied to $S=1$ chains, and we do likewise for $S=1$ junctions. 
$S=1$ chains with open boundary conditions have a localized state with $S_z=1/2$ and localization 
length $\xi=6.03$ \cite{21} at each end. GW \cite{17} obtained four localized $S_z=1/2$ states in 
a Y junction with $N=181$, one at the end of each arm and one centered on the junction.\\

\begin{figure}[t]
\begin{center}
\includegraphics[width=0.5\textwidth]{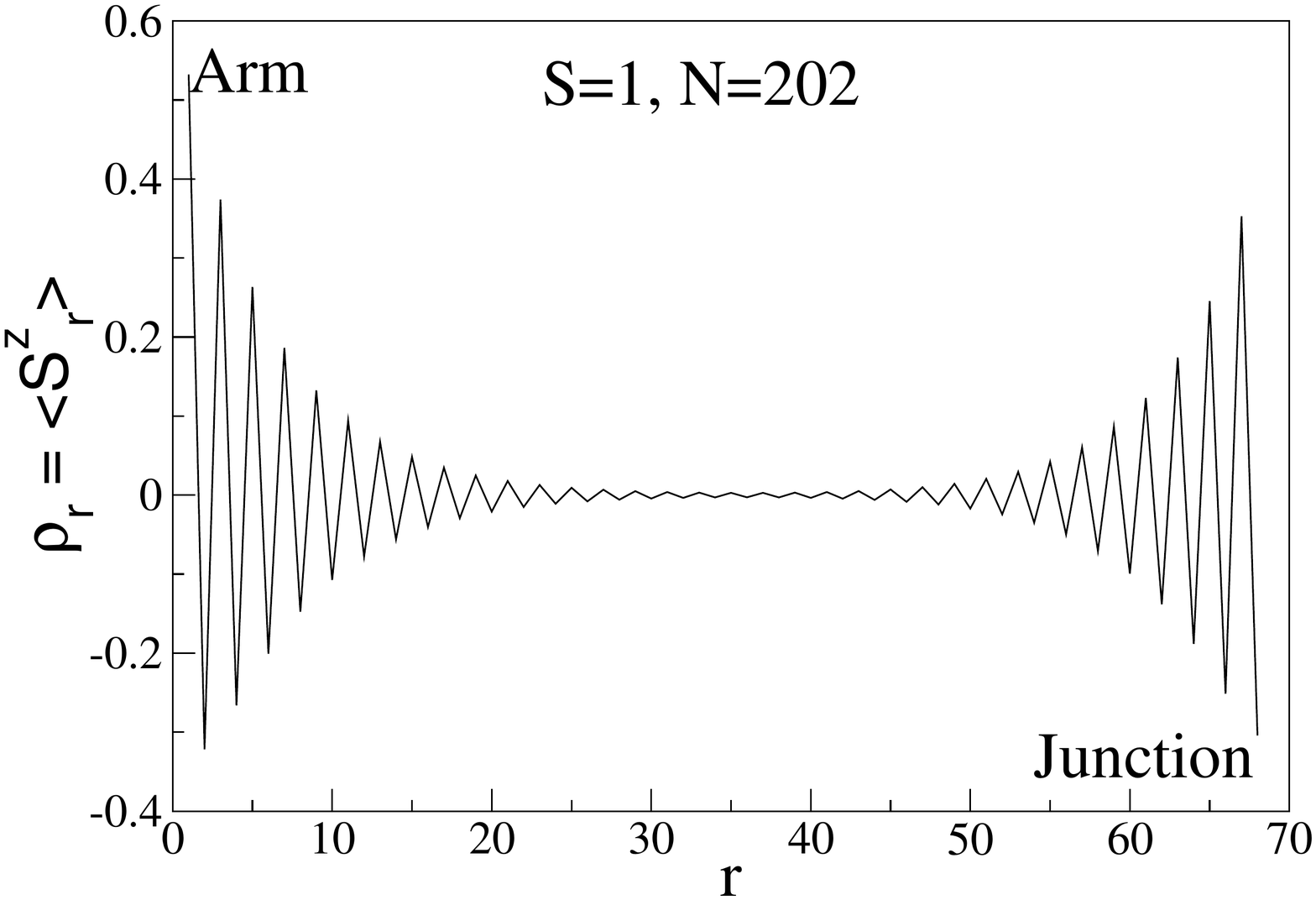}
\caption{Spin densities in one arm of a Y junction of $N=202$ sites, $S=1$, as a function of $r$ with 
$r=1$ at the first site and at $r=68$ at the junction.} 
\label{Fig6}
\end{center}
\end{figure}

\begin{figure}[b]
\begin{center}
\includegraphics[width=0.5\textwidth]{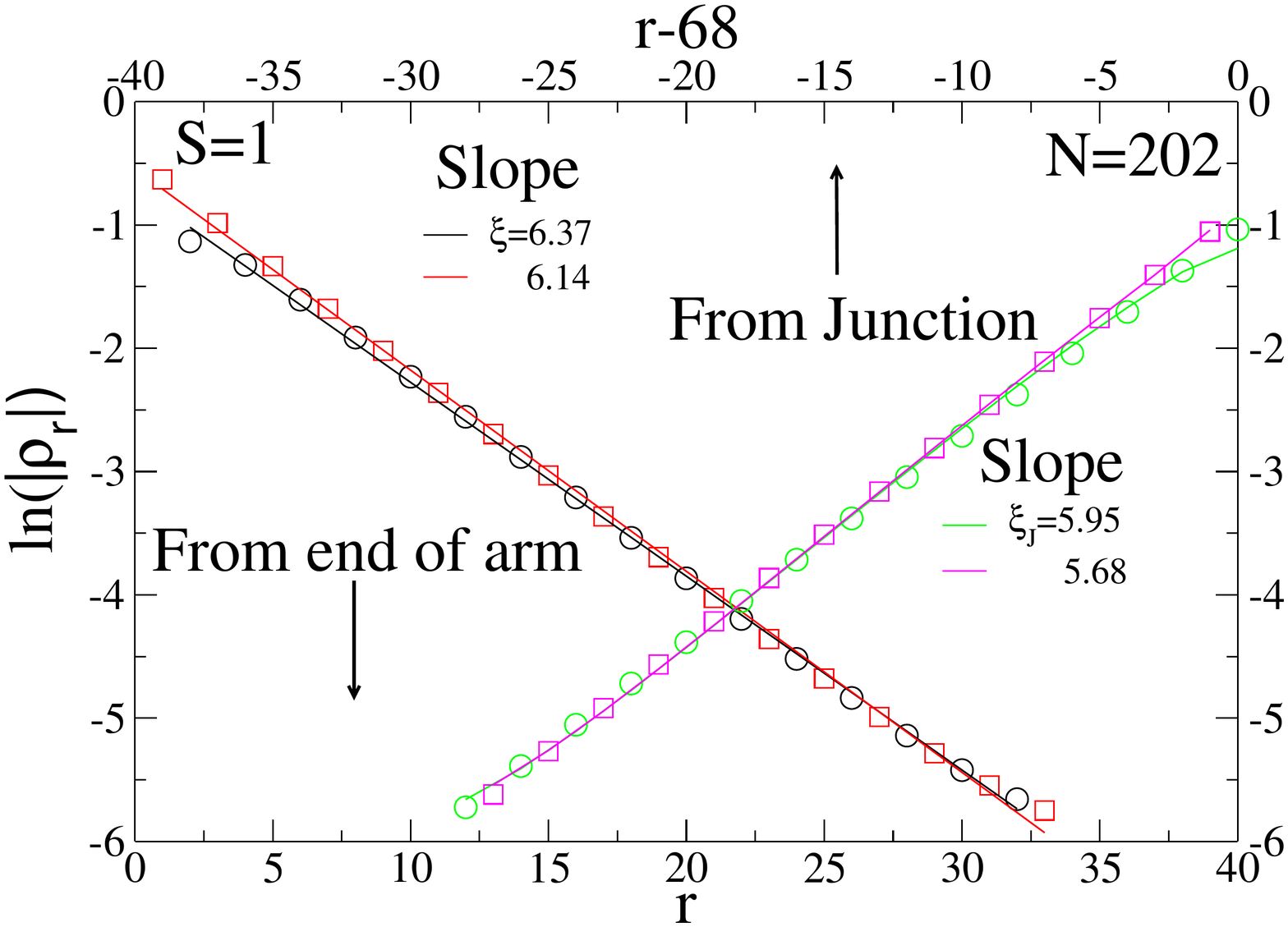}
\caption{Spin densities from Fig. \ref{Fig6} plotted as $ln{\vert{\rho_{r}}}\vert$ vs. $r$
at an arm and at the junction. Circles (squares) represent negative (positive) spin density.} 
\label{Fig7}
\end{center}
\end{figure}

Figure \ref{Fig6} shows the spin densities in one arm of a Y junction of $N=202$ sites with $S=1$. 
As expected, the GS has $S_G=2$, the junction at $r=68$ has $\rho_{J} < 0$, and the total $S_z$ of 
either localized state is 1/2. The spin densities in the first 15 sites of an arm and 14 sites from 
the junction are listed in Table \ref{table3}. The spin densities of localized states are conventionally 
taken as proportional to \cite{21,28}

\begin{equation}
{\rho_{r}} \propto (-1)^{r-1}exp(-r/{\xi})
\label{eq8}
\end{equation}
\noindent
where $r=1$ refers to the ends of chains. This approximation neglects the difference between 
$\rho_{r}$ for even and odd $r$ that is clearly seen in Fig. \ref{Fig6} and Table \ref{table3}. 
Any pair $r$, $r+2$ defines a local localization length \cite{28} 
$\xi=2/(ln{\vert{\rho_{r}}\vert}-ln{\vert{\rho_{r+2}}\vert})$. 
As seen in Fig. \ref{Fig7}, the $\rho_{r} > 0$ and $\rho_{r} < 0$ series have similar 
localization whose average is $\xi=6.25$ for arms, $\xi_{J}=5.81$ for the junction, and the first 
few sites deviate from a simple exponential. White and Huse \cite{21} obtained $\xi=6.03$ for $S=1$ 
chains with open boundary conditions; they did not consider positive and negative spin densities 
separately. GW \cite{17} report similar localization at the junction and arms without going into 
detail, while we find slightly but distinctly smaller $\xi_{J}=5.81$.\\

We consider next Y junctions of $S=1$ spins and odd $N=199$. The GS is a triplet, $S_G=S_z=1$, and 
the junction has $\rho_{J} > 0$. Quite remarkably, the spin densities of the localized state are 
identical to a part per $10^{4}$ to the $N=202$ values in Table \ref{table3} aside from a reversed sign 
around the junction. The localization lengths $\xi_{A}=6.25$ and $\xi_{J}=5.81$ obtained for $N=202$ 
are equally applicable to $N=199$ within our numerical accuracy. Spin densities near the junction 
add to three localized states at the ends of arms for $N=202$ and $S_z=2$, while they subtracts for 
$N=199$ and $S_z=1$. Identical $\vert{\rho_{r}}\vert$ for $N=199$ and 202 directly confirm that each 
localized state has $S_z=1/2$. The GS of a Y junction of $S=1$ spins and long arms is $2^{4}=16$-fold 
degenerate and comprises a quintet, three triplets and two singlets. The quintet has A symmetry under 
$C_{3}$, the singlets transform as E, and the triplets as A and E. 

\begin{table}[t]
\centering
\caption{Spin densities of a Y junction of $N=202$ sites with $S=1$. Listed are the first 15 
sites of an arm, the junction and up to 14 sites from the junction.  }

\begin{tabular}{|>{\centering\arraybackslash} m{8em} |>{\centering\arraybackslash} m{8em} |>{\centering\arraybackslash} m{8em} |} 
\hline
spindensity$\rho_{r}$ & Arm, $r=1$ & Junction \\ 
\hline
1 &     0.5321  & -0.3044 \\[1ex]
2 &    -0.3209  &  0.3530\\[1ex]
3 &     0.3733  & -0.2515\\[1ex]
4 &    -0.2652  &  0.2459\\[1ex]
5 &     0.2624  & -0.1886\\[1ex]
6 &    -0.2000  &  0.1737\\[1ex]
7 &     0.1855  & -0.1383\\[1ex]
8 &    -0.1469  &  0.1234\\[1ex]
9 &     0.1317  & -0.1004\\[1ex]
10 &   -0.1068  &  0.0880\\[1ex]
11 &    0.0939  & -0.0726\\[1ex]
12 &   -0.0773  &  0.0629\\[1ex]
13 &    0.0671  & -0.0629\\[1ex]
14 &   -0.0558  &  0.0450\\[1ex]
15 &    0.0480  & -0.0377\\[1ex]
\hline
\end{tabular}
\label{table3}
\end{table}

\subsection{Junctions with $S>1$}
The Haldane gap of the infinite spin-2 chain is smaller, \cite{29} $\Delta{(2)}=0.0886 \pm 0.0018$, 
about $\Delta{(1)}/5$ and is less accurately known than $\Delta{(1)}$. The ends of $S=2$ chains are 
expected to have localized $S=1$ states with correspondingly larger $\xi$. Schollw\"{o}ck {\it{et al.}} \cite{28} 
have discussed the $S=2$ chain in detail using DMRG, quantum Monte Carlo and exact diagonalization methods; 
they interpret results in a VBS framework and report \cite{30,28} limited agreement. Chains of $N=270$ spins
 with increasing $m$ (to 180) cover a 25-fold spin density change.  Fig. \ref{Fig6} of ref. 28 shows 
$\xi=2/(ln{\vert{\rho_{r}}\vert}-ln{\vert{\rho_{r+2}}\vert})$ to vary as $\sim 40 \pm 10$ up to $r \sim 30$ 
and to be almost constant, $\xi \sim 50$, in the range $40 < r < 125$.\\

\begin{figure}[t]
\begin{center}
\includegraphics[width=0.5\textwidth]{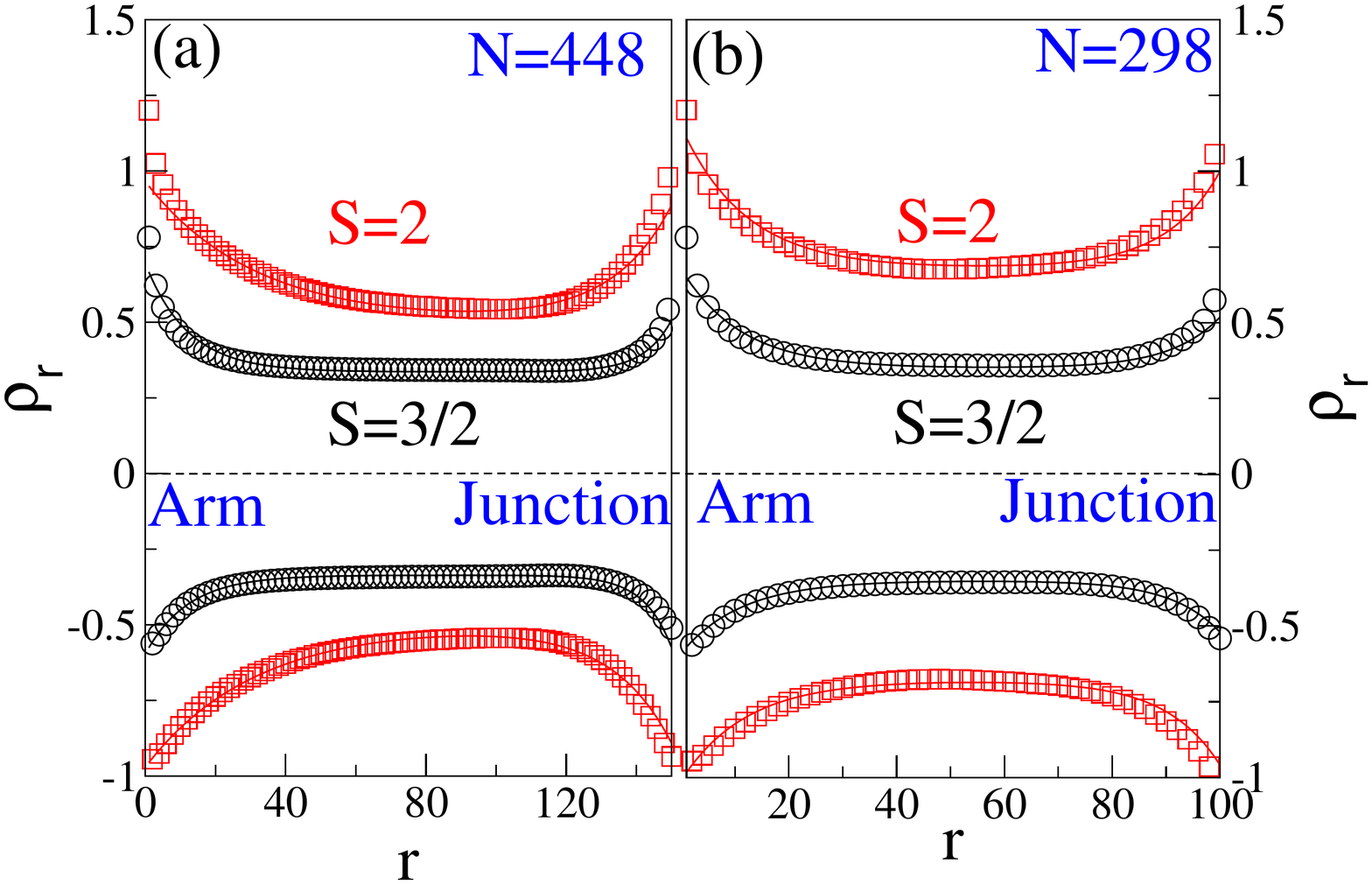}
\caption{Spin densities $\rho_{r}$ in one arm of Y junctions with (a) $N=448$ and (b) $N=298$ spins $S=3/2$ and 2. Lines are Eq. \ref{eq9} with parameters in Table \ref{table3}}
\label{Fig8}
\end{center}
\end{figure}

Smaller $\Delta{(2)}$ and localization $\xi \sim 50$ in $S=2$ chains indicate that Y junctions with longer 
arms are needed to study localized $S=1$ states. Instead of 
localized states, however, and in sharp contrast to $S=1/2$ or 1 junctions, we find substantial AF order 
in both $S=2$ and $S=3/2$ junctions as shown in Fig. \ref{Fig8} for $N=448$ (left panel) and 298 (right panel). 
The junction is at $r=150$ or 100, respectively. The spin density in the interior of arms oscillates between 
$\pm c$ at odd and even $r$. The amplitude increases at the junction and at the end of arms, in contrast to 
the spin densities of the $S=1/2$ junction in Fig. \ref{Fig5} whose magnitude decreases from the junction. 
The similarity in the behavior of the $S=3/2$ and $S=2$ junctions is noteworthy since the infinite $S=3/2$ chain 
is gapless unlike the $S=2$ chain. We consider the main features together before pointing out differences 
between $S=3/2$ and 2 junctions.\\

By definition, Heisenberg exchange is between localized spins $S$ at every site. The sum over $\rho_{r}$ is 
the z component of spin in the given state. The sum over $\vert{\rho_{r}}\vert$ normalized to $NS$ is the 
fraction of unpaired spins; the N\'{e}el state $\vert{AF}\rangle$ with $\pm{S}$ returns 
${(NS)^{-1}}{\sum_{r}{\vert{\rho_{r}}\vert}}=1$. We interpret SDW amplitudes $c$ in Fig. \ref{Fig8} in the 
interior of arms as AF order $c/S$ that increases with $S$. The fraction of unpaired spins in $S=3/2$ 
junctions is 0.293 for $N=245$ and 0.302 for $N=448$; the fraction for $S=2$ spins is 0.326 for $N=445$ 
and 0.323 for $N=448$. By contrast, the fraction is less than $0.1$ in $S=1$ junctions for $N=202$ or 199 
and clearly vanishes in the infinite junction since unpaired spins are in localized states. The fraction 
of unpaired spins also goes to zero in $S=1/2$ junctions with increasing $N$ as discussed earlier.\\  

We generalize Eq. \ref{eq8} for $\rho_{r}$ to reflect the different behavior of spin densities near 
the junction and ends of arms. We study positive and negative $\rho_{r}$ separately but use the same 
length for simplicity, $\xi$ for arms and $\xi_{J}$ for the junction. The spin densities in one arm 
run from $r=1$ to $n$, with $N=3n+1$ and the junction at $r=n+1$. For even $N$ (odd $n$), we analyze 
the spin densities according to 

\begin{equation}
\begin{aligned}
{\rho_{2r-1}} &=c+aexp(-(2r-1)/{\xi})+{a_{J}}exp(-(n+2-2r)/{\xi_{J}})\\
{\rho_{2r}} &=-(c+bexp(-2r/{\xi})+{b_{J}}exp(-(n+1-2r)/{\xi_{J}})).
\end{aligned}
\label{eq9}
\end{equation}

\noindent
The first sum over odd sites has $r= 1, 2,..,(n+1)/2$; the second sum runs to $(n-1)/2$ since 
there is one fewer even site. Both positive and negative spin densities for odd $N$ (even $n$) 
have $r$ up to $n/2$ in Eq. \ref{eq9}.\\

\begin{figure}[b]
\begin{center}
\includegraphics[width=0.5\textwidth]{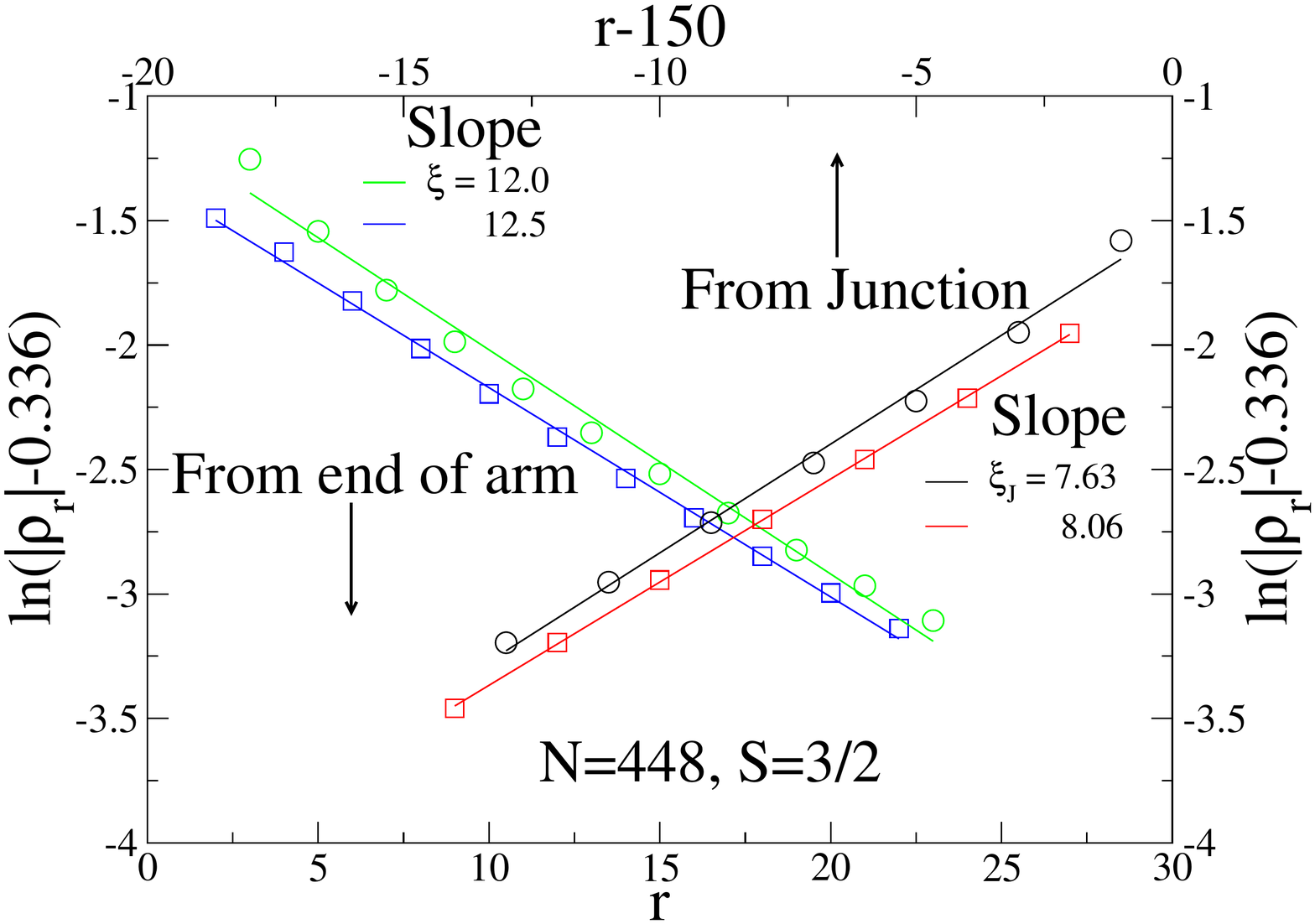}
\caption{Spin densities from Fig. \ref{Fig8} for $S=3/2$, $N=448$ plotted as 
$ln({\vert{\rho_{r}}\vert}-0.336)$ vs. $r$ at an arm and at the junction.} 
\label{Fig9}
\end{center}
\end{figure}

\begin{figure}[t]
\begin{center}
\includegraphics[width=0.5\textwidth]{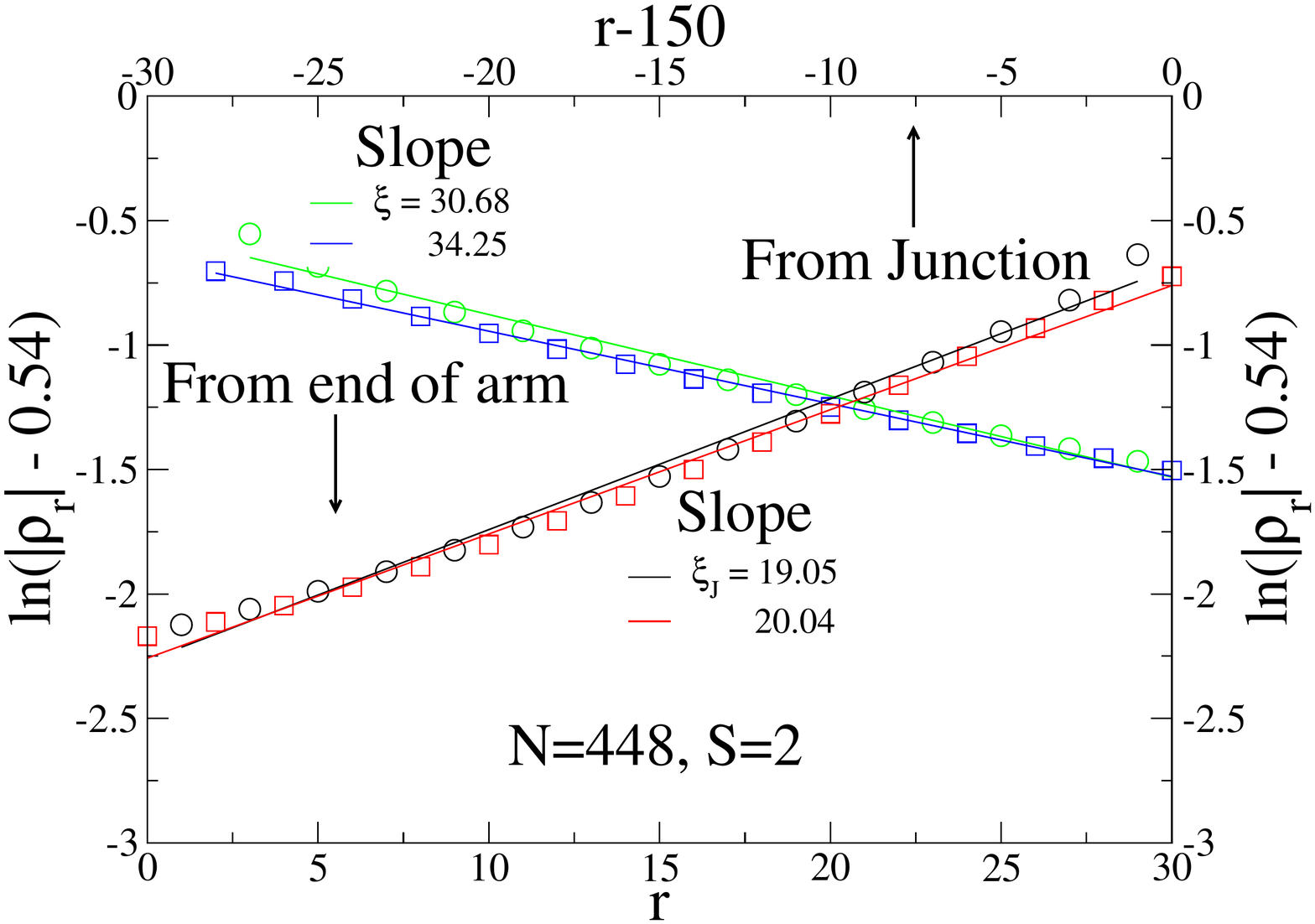}
\caption{Spin densities from Fig. \ref{Fig8} for $S=2$, $N=448$ plotted as 
$ln({\vert{\rho_{r}}\vert}-0.54)$ vs. $r$ at an arm and at the junction.} 
\label{Fig10}
\end{center}
\end{figure}

Exponential contributions are limited to either end when $n > {\xi}$ or $\xi_{J}$. The parameters 
are obtained as shown in Fig. \ref{Fig9} for $N=448$ $(n=149)$ and $S=3/2$. The SDW amplitude is 
$c=0.336$, $\xi_{J}$ is significantly smaller than $\xi$ and ${\pm} {\rho_{r}}$ lead to nearly 
equal $\xi$ or $\xi_{J}$. The junction has $\rho_{J}=-0.510$ for $N=448$ and 0.480 for $N=445$; 
the arms have equal spin densities within a percent or two. The $N=445$ junction has essentially 
the same parameters in Eq. \ref{eq9}. Nevertheless, we have $S_G=3$ for $N=448$, $S_G=3/2$ for $N=445$. 
The difference is largely due to the junction and its first few neighbors. Both Y junctions with even 
or odd $N$ support a SDW with equal $\rho_{1} > 0$ at the end of all three arms. Figure \ref{Fig10} 
shows exponential contributions for $S=2$ junctions with $N=448$ and $S_G=4$. Since the slow decrease 
of spin densities in Fig. \ref{Fig8} and the resulting $\xi \sim 32$ are in the expected range for 
$\Delta{(2)} \sim \Delta{(1)}/5$, the SDW amplitude of $S=2$ junctions may not be entirely due to end 
effects. We leave open the behavior of much longer junctions.\\

Table \ref{table4} lists the parameters of Eq. \ref{eq9} for junctions with even $N$ that generate the 
lines in Fig. \ref{Fig8}; essentially the same parameters hold for $N \pm 1$. The size dependence of the
$S=3/2$ junction has apparently saturated or almost saturated at $N \sim 450$, but has not saturated for 
the $S=2$ junction. We always find $\xi_{J} < \xi$, faster decrease of the SDW amplitude near the junction.\\ 

\begin{table}[t]
\centering
\caption{Eq. \ref{eq9} parameters for Y junctions of $N$ sites with spin $S$} 
\begin{tabular}{|>{\centering\arraybackslash} m{5em} |>{\centering\arraybackslash} m{3em} |>{\centering\arraybackslash} m{3em} |>{\centering\arraybackslash} m{3em} |>{\centering\arraybackslash} m{5em} |>{\centering\arraybackslash} m{5em} |} 
\hline
$S,N$ & $c$ & $\xi$ & $\xi_{J}$ & $a, a_{J}$ & $b, b_{J}$ \\ 
\hline
3/2, 148&  0.433&  5.2&  	5.7&    0.38, 0.25&   0.24, 0.20\\
3/2, 298&  0.350&  11.2&	9.1&	0.30, 0.21&   0.24, 0.19\\
3/2, 448&  0.336&  12.2&	7.9&	0.33, 0.22&   0.26, 0.27\\
2, 298&    0.68&   12&	        10&	0.43, 0.39&   0.33, 0.34\\
2, 448&	   0.54&   32&          20&	0.58, 0.50&   0.52, 0.47\\[1ex]
\hline
\end{tabular}
\label{table4}
\end{table}

The VBS picture has localized $S=1$ states at ends of $S=2$ chains or the arms in $S=2$ junctions. 
We find decreasing $\rho_{1}=1.202$ and $1.200$ for $N=298$ and 448 junctions while Schollw\"{o}ck {\it {et al.}} 
\cite{28} report $1.13$ for a chain $N=270$ with a fixed $S=1$ defect at the other end. The next site has 
$\rho_{2} \sim -0.95$. The first $5-10$ spin densities deviate significantly from Eq. \ref{eq8}, and we 
do not know how to identify a localized state. The $S=3/2$ junction for $N=298$ and 448 has decreasing 
${\rho_{1}=0.781}$ and $0.780$ that, perhaps coincidentally, is again slightly larger than $S/2=0.75$.
Since the SDW amplitude is $S$ in N\'{e}el state $\vert{AF}\rangle$, quantum fluctuations reduce AF 
order by $50\%$ at the ends of arms and by more than $50\%$ elsewhere. SDWs occur naturally in systems 
whose GS has $S_{G} > 0$ and $2S_{G}+1$ degeneracy in $S_z$.\\

To conclude this Subsection, we comment on $S=3/2$ and 2 chains with open boundary conditions that 
were motivated $(i)$ by the unexpected result that $S=2$ junctions do not follow VBS and $(ii)$ to 
confirm quantitative agreement with Schollw\"{o}ck {\it {et al.}} \cite{28}. The GS of quantum chains with 
an even number of spins $N$ is a singlet, $S_{G}=0$. It is not degenerate, thereby excluding a SDW, but 
may have quasi-long-range order in the infinite chain. Delocalized states are expected in the gapless 
$S=3/2$ chain. The gapped $S=2$ chain may have localized $S=1$ states at either end that become decoupled 
in the infinite chain. Two localized states lead to exponentially small gaps between the singlet GS, a 
triplet and a quintet, just as $S=1$ chains have an exponentially small gap to the lowest triplet \cite{31}. 
Accordingly, we studied the quintet, $S_G=2$, with the lowest energy of $S=2$ chains and for comparison 
the lowest-energy triplet, $S_G=1$, of $S=3/2$ chains.\\

\begin{figure}[t]
\begin{center}
\includegraphics[width=0.5\textwidth]{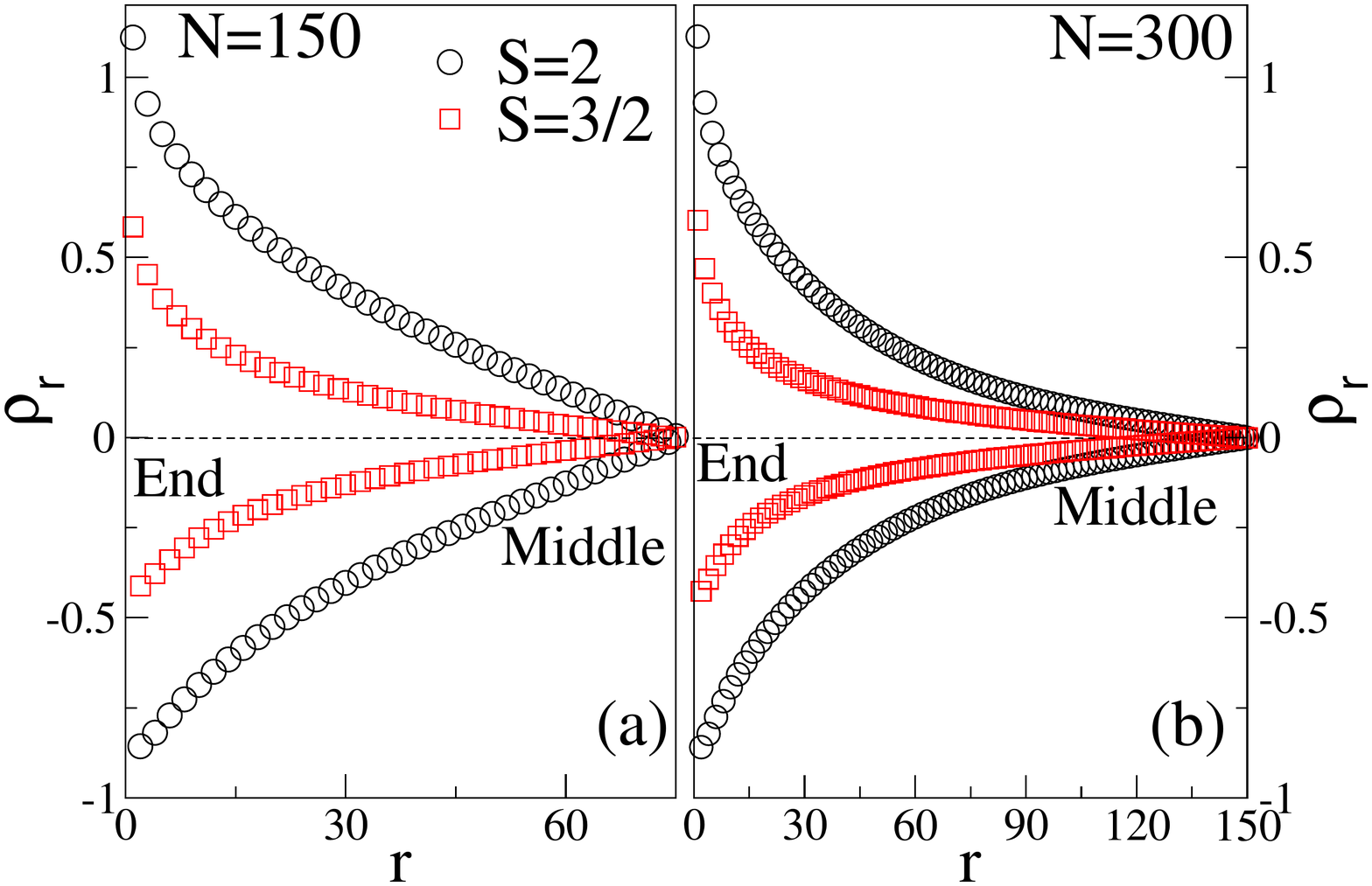}
\caption{Spin densities $\rho_r$ in half chains of $(a)$ $N=150$ and $(b)$ $N=300$ spins 
$S=3/2$ and 2 with antiferromagnetic Heisenberg exchange $J$ between neighbors} 
\label{Fig11}
\end{center}
\end{figure}

Spin densities for open $S=3/2$ and 2 chains of $N=150$ and 300 spins are shown in Fig. \ref{Fig11} up to 
the middle, where they are zero by symmetry. In either case, the first few $\rho_{r}$ depend weakly on size, 
as found previously \cite{28}, and are almost the same as in $N=450$ junctions with 150-site arms. End 
effects are similar in chains and junctions, and exponential fits over a limited range are possible 
aside from the first few spin densities. Symmetry about the middle of chains leads to linear $\rho_{r}$ 
around $r=N/2$ as shown in Fig. \ref{Fig11}. The fraction of unpaired spins is large: 0.183 and 0.121 for 
$S=2$, $N=150$ and 300; $0.090$ and $0.067$ for $S=3/2$, $N=150$ and 300. We infer that the spin densities 
are primarily due to end effects in these chains or junctions, in sharp contrast to the localized states 
in $S=1$ chains or junctions.\\

\section{Discussion}
\label{Sec:IV}
We have presented a modified DMRG algorithm for Y junctions in Section \ref{Sec:II} and results in 
Section \ref{Sec:III} for junctions up to 500 sites, mainly junctions in Eq. \ref{eq1} with Heisenberg 
exchange $J$ between spins $S=1/2$, 1, 3/2 or 2. Much longer chains of, say, 1000 sites greatly increase 
the computational effort at the finite DMRG step. The accuracy may not be lower, however, since the
entanglement entropy \cite{26} of the GS for dividing the junction into system and environmental blocks will
increase only slightly. As already noted, we are considering large but finite junctions rather than
the thermodynamic limit. That limit is better studied in chains since neither the junction nor the ends of arms
should matter in junctions with infinitely long arms.\\

The accuracy of the modified algorithm is fully equal to the DMRG accuracy for 1D chains. Two arms are never
combined into one and new sites are always bonded to the most recently added sites. As in chains, the superblock
Hamiltonian contains only new or once renormalized operators. Infinite DMRG is accurate for $S=1/2$ or 1 junctions,
where finite DMRG makes minimal improvements, but finite DMRG significantly improves the results for $S=3/2$ or 2
junctions. Three or four sweeps of finite DMRG is sufficient for good energy convergence. We performed $5-10$
sweeps for spin densities in order to confirm the different GS of $S=1$ and 2 junctions.\\

The modified algorithm for Y junctions of equal arms can be generalized to other systems, to be discussed
elsewhere \cite{33}. All generalizations are based on the schematic procedures in Figs. \ref{Fig2} and \ref{Fig3}
for equal arms. $(i)$ No change is required for more than three equal arms, although computational requirement
increase as discussed on Section \ref{Sec:II} on going from chains to three arms. $(ii)$ The algorithm performs well in
preliminary tests of Y junctions with arms of different lengths $n \ne n' \ne n''$ \cite{33}. The infinite algorithm
with equal arms is run until the longest arm $n$ is reached. Finite DMRG is then done using blocks of different
size to construct the superblock. $(iii)$ GW considered Y junctions with arms that meet at an equilateral
triangle instead of a point \cite{17}. For such systems, the modified infinite algorithm can again be used to
generate the desired junction. In the beginning of finite DMRG, the superblock is constructed using blocks of
different size; \cite{33} two blocks have the same size, the third block has one fewer site, and the new site is
added to the third block. The modified algorithm can also be generalized to $(iv)$ Y junctions with different
$S$; we use four new sites with different $S$ and three arms at every step \cite{33}. A new site is added at the 
end of each arm and another one is added at the junction of these three arms in Fig. \ref{Fig2}. Since the size 
of the density matrix is $L=(2S+1)^{2}m^{2}$ for adding one spin $S$, the size scales as $(2S+1)^{8}$ for adding 
four spins $S$. The procedure is efficient for small $S$ but rapidly becomes more expensive for large $S$.\\

Kekul\'{e} diagrams are special cases of VB diagrams in which singlet pairing is limited to adjacent atoms or to
adjacent $S=1/2$ sites. The complete VB basis also has singlet pairing of more distant sites. There are $N_A$
Kekul\'{e} diagrams for the $S=1/2$ junction with odd $N$: the unpaired electron is at a site of the larger
sublattice, which uniquely fixes singlet pairing, shown as lines to neighbors, of all remaining sites. Although
the total number of VB diagrams is exponentially larger, Kekul\'{e} diagrams are often a simple approximation for
fermionic or spin-1/2 systems that provides semi-quantitative information; they have been extensively used in
organic chemistry for well over a century. Their scope now extends to VBS in which $S \geq 1$ sites are represented
\cite{20, 28, 32} as two or more $S=1/2$ spins that are singlet paired with neighbors. VBS  has $2S$ lines
to neighbors and no unpaired spins aside from chain ends. Unpaired spins and resonance among Kekul\'{e} diagrams account
naturally for localized $S_{z}=1/2$ state in $S=1$ junctions, albeit without any reference to the Haldane gap.\\

VBS wave functions are exact GS of special Hamiltonians that Schollw\"{o}ck {\it{et al.}} \cite{28} discuss and write
explicitly for $S=2$ chains. Heisenberg chains with finite $\Delta{(S)}$ for integer $S$ have bilinear exchange
$J S_{r} \cdot S_{r+1}$ between neighbors while $H_{VBS}$ contains terms up to $(S_{r}\cdot S_{r+1})^{4}$ for $S=2$.
Schollw\"{o}ck and Jolicoeur \cite{30} find that the Haldane phase described by VBS is strongly reduced in
$S=2$ chains compared to $S=1$ chains. Our DMRG results for finite Y junctions are quite consistent with VBS for
$S=1$ junctions where quantum fluctuations suppress AF order. DMRG for $S=2$ junctions does not follow VBS,
however. The GS is instead a SDW with reduced but finite AF order.\\

\textbf{Acknowledgements} MK thanks DST for a Ramanujan Fellowship SR/S2/RJN-69/2012 
and DST for funding computation facility through SNB/MK/14-15/137. ZGS thanks NSF for 
partial support of this work through the Princeton MRSEC (DMR-0819860). SR thanks DST 
India for financial support.

\end{document}